# Energy Transfer in Stability-Optimized Perovskite Nanocrystals

*Michèle G. Greiner[‡], Andreas Singldinger[‡,]\*, Nina A. Henke, Carola Lampe, Ulrich Leo, Moritz Gramlich, and Alexander S. Urban\**

Nanospectroscopy Group and Center for Nanoscience (CeNS), Nano-Institute Munich, Department of Physics, Ludwig-Maximilians-Universität München,

Königinstr. 10, 80539 Munich, Germany

ABSTRACT

Outstanding optoelectronic properties and a facile synthesis render halide perovskite nanocrystals (NCs) a promising material for nanostructure-based devices. However, the commercialization is hindered mainly by the lack of NC stability under ambient conditions and inefficient charge carrier injection. Here, we investigate solutions to both problems, employing methylammonium lead bromide (MAPbBr$_3$) NCs encapsulated in diblock copolymer core-shell micelles of tunable size. We confirm that the shell does not prohibit energy transfer, as FRET efficiencies between these NCs and 2D CsPbBr$_3$ nanoplatelets (NPLs) reach 73.6%. This value strongly correlates to the micelle size, with thicker shells displaying significantly reduced FRET efficiencies. Those high efficiencies come with a price, as the thinnest shells protect the encapsulated NCs less from environmentally induced degradation. Finding the sweet spot between efficiency and protection could lead to the realization of tailored energy funnels with enhanced carrier densities for high-power perovskite NC-based optoelectronic applications.

INTRODUCTION

Since their first emergence in 2014, halide perovskite nanocrystals (NCs) have been drawing growing interest due to their exciting fundamental properties and their immense commercial potential.[1,2] Scientifically, their flexible compositions and precise size control allow for studies on exciton fine structure, level inversion, polaron formation, carrier dynamics, and phonon interaction.[3-9] With emission wavelengths tunable throughout the visible range, quantum yields approaching unity, cheap and facile syntheses, and abundant precursor materials, potential applications range from light emission (light-emitting diodes (LEDs), lasers, and displays) to solar cells, photodetectors, field-effect transistors, and even photocatalysis.[10-14] Despite sounding like the perfect material, halide perovskites also (currently) exhibit limitations that have been impeding their widespread commercialization. Perhaps the most critical property in this regard is stability, as incorporated NCs need to survive for the intended lifespan of the device. Unfortunately, halide perovskites have inherent susceptibility to extrinsically-induced degradation through moisture, UV-light, and heat, as well as a large mobility of (halide) ions.[15-17] Strategies to mitigate environmentally-induced degradation initially focused on embedding the NCs inside stable matrices, such as polymer coatings, solid and mesoporous silica, or metal-organic frameworks (MOFs).[18-20] A drawback of these strategies is that NCs are often not well dispersed or matrix grains are very uneven in size, reducing the quality of active layer films and concomitantly device performance. Moreover, NCs cannot be investigated individually for scientific purposes.

More recently, research groups have been concentrating on creating core-shell NCs, embedding individual perovskite NCs in nanometric silica, alumina, or polymer shells.[21-23] These heterostructures have shown significantly enhanced stability while retaining the optical properties of the perovskite NCs. Our previous work was based upon using diblock copolymers as nanoreactors for the perovskite synthesis and resulted in encapsulated perovskite NCs that could survive full water submersion for more than 75 days. Importantly, this approach can be conducted in ambient conditions without complex apparatuses or preparatory steps. While the nonconductive polymer shielding could be expected to preclude optoelectronic integration, we showed that Förster resonance energy transfer (FRET) occurs between layers of subsequently deposited bromide- and iodide-based NCs with a FRET efficiency $\eta_{FRET} \approx 31\%$. Obviously, this value is a result of the large separation between the NCs and is too low to yield high-performance devices. Accordingly, the question is how much polymer shielding is necessary to sufficiently stabilize the NCs while retaining enough energy transfer capability to fabricate efficient optoelectronic devices.

To this end, in this work, we synthesized block copolymer-encapsulated NCs with varying polymer lengths and, accordingly, micelle sizes to investigate the trade-off between stability and energy transfer in NC-heterostructure mixtures. To avoid issues of halide ion exchange, we employed all bromide-NCs with encapsulated $MAPbBr_3$ NCs acting as energy acceptors and two-dimensional $CsPbBr_3$ NPLs as donors. All encapsulated NCs exhibited nearly identical optical properties regardless of the polymer used. Energy transfer efficiencies in the hybrid system of up to 73.6% are reached using the shortest polymer. Due to a larger inter-NC distance, the FRET efficiency drops by 45% for the worst performing micellar-NC. Thin films of encapsulated NCs were dropcasted on glass substrates, and the development of their photoluminescence (PL) emission was monitored when stored in ambient conditions. The highest performing polymer in terms of energy transfer displayed the shortest stability at 17.6 days. This value rose to 27.5 days for the most stable micellar-NC employing a far longer block copolymer. These results indicate the trade-off between stability and energy transfer efficiency in such a system. Importantly, this trade-off can be controlled and optimized for specific applications and provides a viable strategy to bypass commercialization hurdles such as instability and charge injection inefficiency.

RESULTS

We investigated acceptor NCs fabricated according to our previously reported diblock copolymer nanoreactor synthesis (Figure 1a).[23] We employed the hybrid organic/inorganic $MAPbBr_3$ perovskite and diblock copolymers comprising a combination of polystyrene (PS) and poly(2-vinyl pyridine) (P2VP). In toluene, the P2VP block forms the core, while the PS part constitutes the shell of micelles (PS-b-P2VP) due to the differing solubilities of the two blocks. A subsequent two-step addition of the perovskite precursors leads to a crystallization of halide perovskite within the micelles. We chose five polymer combinations with similar P2VP monomer numbers but PS blocks ranging from 113 to 558 monomers. Accordingly, we name the polymer and the resulting NCs based on this length, e.g., PS113. A detailed list of polymers used is shown in the Supporting Information (see Table S2). Our previous publication determined the core size to depend on the number of P2VP units comprising the micellar cores. Consequently, we did not expect a significant variation in the core sizes for the chosen polymers. We acquired TEM images and found a solid linear correlation of core size with P2VP block length for the three shortest polymers from $8.7 \pm 1.9\ nm$ to $12.4 \pm 3.2\ nm$ (see SI Figures S1 & S2). The two longest polymers (PS394 and PS558) exhibited the smallest core radii of $6.9 \pm 1.9\ nm$ and $7.5 \pm 1.5\ nm$, respectively. Two effects could induce the smaller NC sizes. Firstly, longer shell polymer blocks compress the core polymer block, leading to smaller cores.[24] Secondly, the extremely long polymers likely impede the diffusion of precursor salts into the cores, limiting the speed of NC formation or even the overall core size. To estimate the shell thickness, we acquired AFM images on dip-coated films of the encapsulated NCs (Figure 1b). These revealed highly monodisperse NC monolayers, from which we can extract the core-to-core separation via fast Fourier transform (FFT) for each polymer type (see Figures S3 & S4). Using this distance and the previously determined core sizes, we can estimate the thickness of the polymer shell, which amounts to between $3.7 \pm 1.8\ nm$ (PS113) and $9.5 \pm 2.2\ nm$ (PS394). A sketch of the individual samples is also presented in Figure 1b. Importantly, the encapsulated perovskite NCs show only small deviations in their UV-Vis and PL spectra, with small Stokes shifts, PL emission maxima between $524\ nm$ and $525\ nm$ for the first three micelle species and narrow full width at half maxima (FWHM) of around $22 \pm 1\ nm$ ($100 \pm 3\ meV$). The latter two of the five micelle species show slightly blue-shifted PL emission maxima at 523.4 nm and 520.3 nm for the PS394 and the PS558 micelle, respectively. We assume this originates from weak quantum confinement effects due to hampered precursor salt diffusion and less space in the compressed micelle cores. Several individual crystallization centers could lead to smaller, separated NCs, thereby blue-shifting the overall PL emission (Figure 1c, Figure S5, see Table S3 for full data).

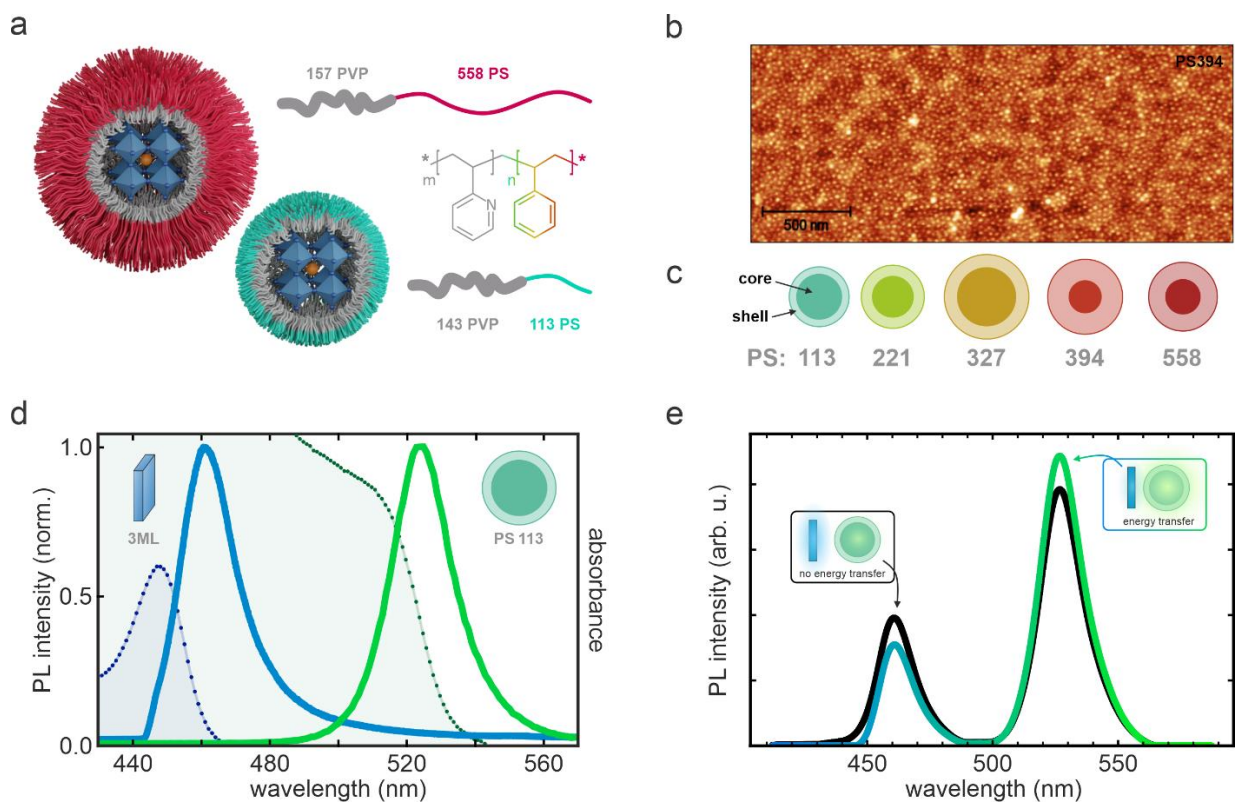

**Figure 1. a)** Schematic of a micelle-encapsulated perovskite NC with the thickest shell (PS558) and one with the thinnest shell (PS113) with their corresponding diblock copolymer. **b)** AFM image of a film of dropcasted PS394 micelles, self-arranging in a hexagonal pattern. **c)** Schematic of all five micelle systems with relative core size and shell thickness. **d)** Absorption (dashed lines) and PL spectra (solid lines) of 3ML NPLs and PS113 micelles. **e)** Experimental PL spectrum of a 3ML NPL - PS113 micelle thin film (colored line) and a calculated PL spectrum (black curve) obtained through a weighted addition of the two individual PL spectra. The decrease(increase) in PL emission of the donor(acceptor) indicates an energy transfer process between the two NC species.

We decided to employ the encapsulated NCs as the acceptor in a FRET system to test their optoelectronic integrability. As donor NCs, we chose $Cs_{n-1}Pb_nBr_{3n+1}$ NPLs of 2ML and 3ML thickness and synthesized these according to slightly modified, previously described methods.[5, 25] The main difference in the synthesis was that we did not add the enhancement solution to ensure the photoluminescence (PL) intensity of the two types of NCs were roughly equivalent. The NPLs have lateral sizes of approximately $20x20\ nm$ and a thickness of 1.2 and $1.8\ nm$, respectively (see Figure S6). The strong confinement in one dimension leads to a strong blueshift of the absorption onset and PL emission. Therefore, the spectra can be used to determine the quality of the samples, which are shown to be of high homogeneity (see Figure 1c, Figure S5). Noticeably, the exciton peak is strongly pronounced for the NPLs, yet hardly visible for the micellar-NCs. This indicates that the encapsulated NCs, comparable with previous results, have a size larger than the bulk exciton Bohr radius of 3-5 $nm$ and can therefore be sorted into the weak confinement regime.[26] The PL spectrum of the 3ML NPLs is centered at 455 $nm$ with a FWHM of 19 $nm$ (114 $meV$) while that of the 2ML NPLs is located at 434 $nm$ with a FWHM of 11 $nm$ (72 meV). We chose these NPLs to guarantee a significant overlap of the PL with the absorption of the encapsulated NCs while ensuring enough spacing between the PL maxima to separate the PL signals for the subsequent analysis efficiently. All measurements presented in the main manuscript were carried out with the 3ML NPLs. To confirm the conclusions and compare the results, we repeated the same measurements with the 2ML NPLs (for details, see Supporting Information).

FRET is based on a dipole-dipole interaction between a donor (here: NPLs) and an acceptor (here: encapsulated NCs). FRET between NCs strongly depends on their spectral overlap, the orientation of the transition dipoles, and the distance between both participants.[27] Often, these factors are incorporated into one value, the so-called Förster distance, which signifies the separation between the donor-acceptor pair at which the FRET efficiency is 50%. As mentioned above, the spectra of our NCs are well-matched and fulfill the criteria for spectral overlap (Figure 1c). As the polymer shell thicknesses are within the typical Förster distance range of 5-10 $nm$,[28] we can manipulate the distance between acceptor and donor via the shell thickness and investigate how the FRET efficiency is affected by the NC separation. To this end, we mixed the NC dispersions in specific ratios over several orders of magnitude (see Supporting Information for details). Subsequently, thin films were created by drop-casting the mixtures onto $SiO_2$-coated Si substrates and investigated with a self-built μ-PL setup. Figure 1e shows the PL emission of a mixture of 3ML NPLs and PS113 micelles (colored curve) excited with a laser wavelength of $\lambda = 413\ nm$. To check for energy transfer, we calculated a spectrum from the pure emissions weighted with the corresponding mixing ratio (black curve) for comparison to the experimental data. The calculated spectrum represents the case where no energy transfer takes place. Comparing the two spectra, one observes that the PL emission of the acceptor is increased, whereas that of the donor is decreased for the mixed sample, the first hint of an energy transfer process. PL emission spectra of specific mixing ratios were obtained for micelles formed out of all five different polymers (see SI Figure S7). Notably, all mixtures follow the same trend of increased acceptor emission and decreased donor emission compared to the calculated curves.

To rule out that this is due to the reabsorption of emitted photons in the thin films, we employed time-correlated single-photon-counting (TCSPC) to confirm a FRET-mediated transfer process. Dichroic edge filters are installed into the system to selectively transmit only donor or only acceptor emission (see SI Figure S8). Figure 2a shows donor PL decay curves of mixtures from 3ML NPLs and PS113 micelles for varying mixing ratios. Noticeably, the PL decay becomes progressively faster with increasing acceptor concentration. An additional decay channel causes these changes as energy is transferred non-radiatively via FRET from the donor to the acceptor. Reabsorption of the emitted photons can be neglected since it would not influence the donor PL lifetime. The FRET process influences both the donor and the acceptor decay, with the latter becoming slower by feeding of the acceptor through the donor. Correspondingly, TCSPC measurements on the acceptor reveal an enhanced PL lifetime for an increasing amount of the donor (see SI Figure S9 f-j), confirming this to be a FRET-based process. TCSPC measurements were also conducted on mixtures with 2ML NPLs and micelles. Since the spectral overlap does

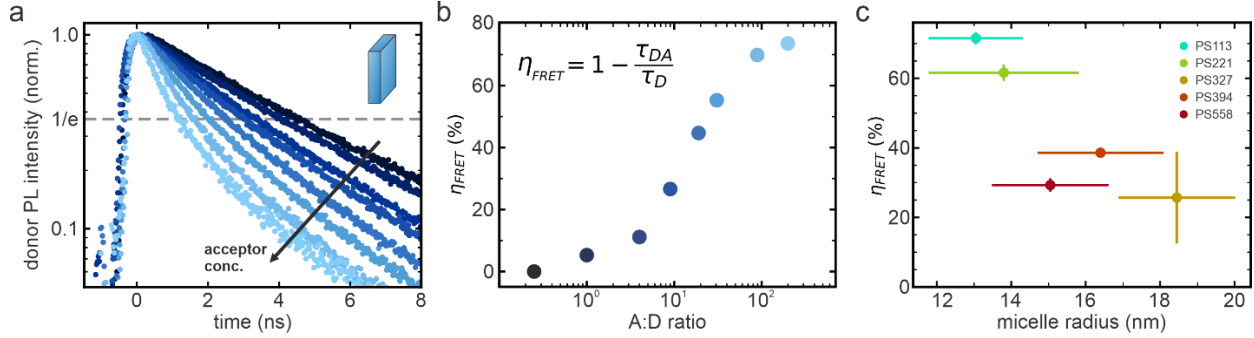

**Figure 2. a)** PL decay curves of a 3ML NPL film (black) and from mixed films with specific A:D ratios (blue). With an increasing A:D ratio, the donor decay becomes progressively faster. **b)** Calculated FRET efficiencies of 3ML and PS113 micelle mixtures as a function of the A:D ratio. **c)** Calculated maximum FRET efficiencies of all five NPL – encapsulated NC systems. The FRET efficiency decreases from 73,6% to 25,7% with increasing micelle radius.

not vary substantially when replacing the 3ML with the 2ML NPLs, the trends of the mixed PL spectra and PL decay curves are the same (see SI Figures S10 and S11). The transfer efficiency is one of the main quantification methods, as was previously reported for CdSe and perovskite NPLs, where values of $\eta_{FRET} = 60 - 70\ \%$ were determined.[29, 30] The transfer efficiency is determined by the change in the donor PL lifetime: $\eta_{FRET} = 1 - \frac{\tau_{DA}}{\tau_D}$

Herein, $\tau_D$ is the lifetime of the pure donor and $\tau_{DA}$ is the PL lifetime of the donor in the donor-acceptor mixture. Here, we must note that the donor lifetime is strongly dependent on the surroundings. As synthesized, the 3ML NPLs display a monoexponential PL decay with a lifetime of $0.9\ ns$, which becomes significantly longer ($2.1–2.3\ ns$) when mixed with empty micelles in toluene (see SI Figure S12) likely through NPL-polymer interactions. Accordingly, this latter value is employed as the pure donor lifetime for all subsequent calculations. Transfer efficiencies calculated for all investigated mixing ratios of the thinnest micelle shell are depicted in Figure 2b. Naturally, with no acceptors present, $\eta_{FRET} = 0$. As the acceptor to donor ratio is increased, $\eta_{FRET}$ is boosted concomitantly as more and more acceptors receive energy from the donor NPLs.

The efficiency saturates at a maximum of nearly 73.6 % once all donors are essentially surrounded by acceptors.

To study the effect of the polymer shielding on the FRET efficiency, we repeated the measurements for all block copolymers several times in the same manner (Figure 2c). The four additional polymers were all longer than the initially studied PS113. Accordingly, these four encapsulated NC samples exhibited lower $\eta_{FRET}$ values down to 25.7% for the PS327 sample. One would expect the FRET efficiency to trend with the shell thickness of the micelle since the shell thickness influences the distance between the participants. However, the FRET efficiency displays a strong correlation with the total radius of the encapsulated NCs instead. We assume this is because an electron-hole pair residing inside a NPL will find the highest density of states for FRET transfer near the center of the encapsulated acceptor NC. Accordingly, the micelle radius is the relevant distance for determining energy transfer efficiency. The observed strong correlation fits the theory of FRET, where an increased distance between the participants reduces the transfer efficiency. The micelle radii given are larger than the standard Förster distances. To understand this, one must consider that the NCs are quite large compared to fluorescent molecules, and excitonic states are located at the edge of the micellar cores. Thus, the relevant separation between NPLs and encapsulated NCs is likely well within the range of Förster distances. The critical distance for energy transfer is, therefore, given by the shell thickness plus the length of the ligand shell surrounding the NPLs, amounting to roughly 4-12 $nm$. This explains why we can still observe FRET for the thickest micelle shells. Since $\eta_{FRET} = 73.6\ \%$ was the highest observed value we can conclude that for this NC system, the effective Förster distance must be on the order of 7-9 $nm$. To confirm that FRET was responsible for the observed changes in the PL decay times, we shifted the laser excitation to $\lambda = 485\ nm$ to only excite the micelles in the mixture (see SI Figure S13). TCSPC measurements reveal that PL lifetimes of micelles in mixtures do not change compared to the pure micelle samples, confirming the previous assertions. To verify the generality of the findings and the FRET process, we repeated the measurements using thinner NPLs (2 ML) and observed similar trends and efficiency values for the different NPL-micelle systems (see SI Figure S14).

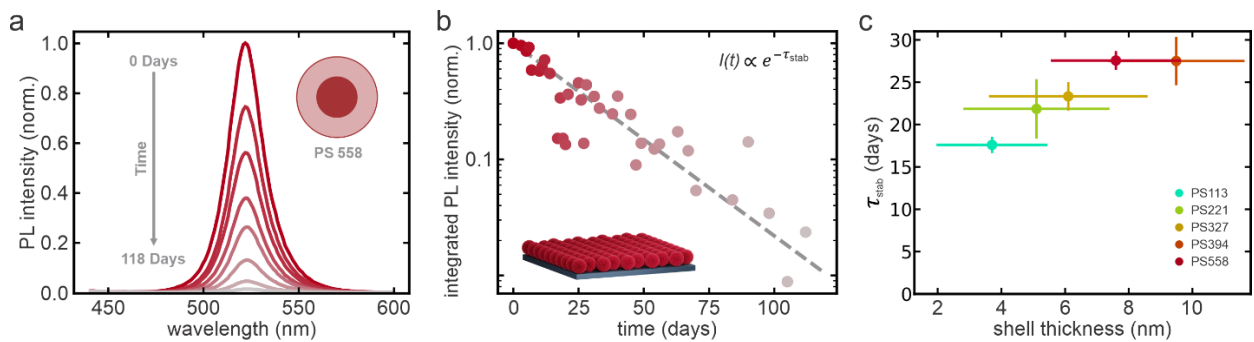

**Figure 3. a)** PL spectra of a dropcasted PS558 micelle film acquired over 118 days. While the overall intensity decreases, the peak position stays constant, indicating no change in the NCs inside the micelle. **b)** Integrated PL intensity of the spectra shown in a) as a function of time. As indicated, the data is fitted with a monoexponential function (grey dashed line). For comparison of the different systems, a stability lifetime $\tau_{stab}$ is extracted from the fit. **c)** Stability lifetime values for each of the five investigated encapsulated NC samples. The thicker the micelle shell, the higher the stability.

The original main goal for employing polymer encapsulation was to enhance the stability of the perovskite NCs. Having shown that by tuning the micelle shell thickness, it is possible to alter FRET efficiencies, we next focused on the stability of the different encapsulated NCs. We prepared drop-casts (50 µl) on glass substrates for each micelle type for the stability measurements and stored them in the laser lab under ambient conditions (approximately 40% relative humidity, 22 °C, and nearly no light) for over 100 days (Figure 3a). Over time, the PL intensity decreases;

however, the PL spectrum retains its shape and spectral position nearly perfectly. Accordingly, even after 118 days, we can still obtain an appreciable PL signal, far longer than for typical halide perovskite NCs.[15, 23] To compare the stability of the different micelle sizes, we integrated the total PL signal and normalized it to the value of the sample on day 0 (Figure 3b). The PL intensity displays a monotonous exponential decay for the duration of the experiment. We extract the time constant of this decay and declare it the stability lifetime. The FRET champion NCs (PS113) displayed the shortest stability lifetime at 17.6 days, while the longest lifetime was obtained for the micelles with the thickest shell (PS558) at 27.5 days (Figure 3c, see SI Figure S15 & S16 for all PL stability measurements). The thickness of the micelle shell has a significant impact on the stability of the NCs. However, in contrast to the FRET efficiency, the lifetime correlates strongly with the shell thickness, possibly exhibiting a saturation behavior for a shell thickness of $> 8\ nm$. An attentive reader might have noticed that these stability lifetimes are shorter than those obtained in our previous publication.[23] One must note that we used much less encapsulated NC dispersion to produce the films here. Accordingly, to compare to our previously determined stability lifetimes, we made several more drop-casts using four times the drop-cast volume ($200\ \mu l$) and monitored the PL intensity (see SI Figure S17). The PL intensity of the thicker films comprising the PS113 micelles remained highly stable, showing no discernible drop in PL intensity after 100 days. Even more impressive, the PS558 micelles exhibited increasing PL intensities over the entire period, reaching 110% after 100 days. Clearly, not only the shell of the micelle is important for stability, but the film thickness also, as encapsulated NCs additionally protect other NCs lying beneath them within the films. The increase in PL intensity could result from a gradual crystallization process within the micelles, possibly due to the aggregation of nanocrystallites within the micelles and a reduction in surface traps. This is possible in contrast to conventional perovskite NCs, for which the organic ligands prevent further aggregation. Additionally, we investigated the stability of 50 µ$l$ and 200 µ$l$ dropcasted films at elevated temperature (60°$C$) and under continuous excitation (see SI Figure S18 to S21). For both measurements, we observed a faster degradation compared with samples stored in ambient conditions. For samples stored at 60°$C$ the film thickness has a significant impact on the stability from which we conclude an oxygen-based degradation mechanism. However, the samples stored under continuous excitation show a distinct behavior where the film thickness does not affect the stability, indicating a different degradation mechanism.

In summary, we have investigated the stability and efficiency of the energy transfer between $CsPbBr_3$ NPLs and $MAPbBr_3$ NCs encapsulated in block copolymer micelles with five different shell thicknesses. We find that energy transfer occurs via FRET and reaches a maximum efficiency of $\eta_{FRET} = 73.6\,\%$ for the shortest block copolymer. The FRET efficiency correlates strongly with the radius of the NC-micelle system, dropping to 25.7% for the NCs with the largest radius. In contrast, by increasing the polymer length from PS113 to PS558, which corresponds to a shell thickness increase from $3.7 \pm 1.8\,nm$ to $9.5 \pm 2.2\,nm$, the stability of dropcasted thin films in ambient conditions increased from 17.6 days (thinnest polymer shell) to 27.5 days (thickest polymer shell). Notably, we find that not only the shell thickness itself is important for the stability but the overall film thickness. Using four times the amount for preparing the dropcasted films, we observed that the PL intensity remained stable for the thinnest polymer and even increased over 100 days for the longest polymer used. These results are important for obtaining environmentally stable perovskite NCs and providing valuable device fabrication insights.

ASSOCIATED CONTENT

**Supporting Information**. Materials and methods for the synthesis of the NPLs and the micelles. Method on how the NPL – micelle mixtures were created. TEM and AFM images of all five investigated micelles. Micelle size determination via TEM and AFM. PL spectra and PL decay curves of mixtures of 3ML NPLs with all five investigated micelles. Impact of different pure diblock copolymers on the NPL PL decay time. Confirmation of the results with PL spectra and PL decay curves of 2ML NPLs mixed with micelles of different shell thicknesses. PL decay curves of 2ML NPL-micelle mixtures optically excited at a higher wavelength. FRET efficiencies calculated from mixtures of 2ML NPLs with different micelle types. Two measurement series of film stability of all five investigated micelle types. Impact of film thickness on measured stability of dropcasted micelle thin films. The following files are available free of charge.

AUTHOR INFORMATION

**Corresponding Author**

*Andreas Singldinger, Email: andreas.singldinger@physik.uni-muenchen.de

*Alexander S. Urban, Email: urban@lmu.de

Website: www.nanospec.de

Twitter: @NanospecGroup

**Author Contributions**

The manuscript was written through contributions of all authors. All authors have given approval to the final version of the manuscript. ‡ M.G.G. and A.S. contributed equally to this manuscript.

**Funding Sources**

**Notes**

Any additional relevant notes should be placed here.

ACKNOWLEDGMENT

This project was funded by the European Research Council Horizon 2020 through the ERC Grant Agreement PINNACLE (759744), by the Deutsche Forschungsgemeinsschaft (DFG) under Germany's Excellence Strategy EXC 2089/1-390776260 and by the Bavarian State Ministry of Science, Research and Arts through the grant "Solar Technologies go Hybrid (SolTech)".

TABLE OF CONTENTS GRAPHIC

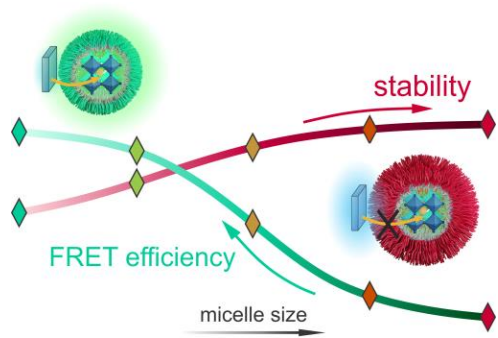